\documentstyle[prd,aps]{revtex}
\begin{document}
\draft
\author{B.V.Ivanov\thanks{%
E-mail address: boyko@inrne.acad.bg}}
\title{Colliding axisymmetric $pp$-waves}
\address{Institute for Nuclear Research and Nuclear Energy\\
Tzarigradsko Shausse 72, Sofia 1784, Bulgaria}
\date{8 May 1997}
\maketitle

\begin{abstract}
An exact solution is found describing the collision of a class of
axisymmetric $pp$-waves. They are impulsive in character and their
coordinate singularities become point curvature singularities at the
boundaries of the interaction region. The solution is conformally flat.
Concrete examples are given, involving an ultrarelativistic black hole
against a burst of pure radiation or two colliding beam-like waves.
\end{abstract}


\section{Introduction}

The problem of colliding plane waves in general relativity has been
thoroughly investigated by now \cite{one}, \cite{two}. Even more interesting
and realistic is the collision of the more general class of $pp$-waves of
finite extent and energy. One particular example, the collision of
ultrarelativistic black holes, has been studied by approximate methods \cite
{three}, \cite{four}, \cite{five}. The main reason for the lack of exact
solutions is that $pp$-waves are written easily in Brinkmann coordinates,
but the analog of the Rosen transformation has not been known. Recently, the
diagonalization of axisymmetric $pp$-waves was achieved \cite{six}. They are
described by the line element in cylindrical coordinates 
\begin{equation}
ds^2=2dudv-e^{-U}\left( e^Vdr^2+e^{-V}d\varphi ^2\right)  \label{one}
\end{equation}
where $u=\frac 1{\sqrt{2}}\left( t-z\right) $, $v=\frac 1{\sqrt{2}}\left(
t+z\right) $ and $U$, $V$ depend on $u,$ $r$ for a left-coming wave and on $%
v,$ $r$ for a right-coming wave.

The standard description of a head-on collision of two waves divides the $u,$
$v$ space into four regions \cite{one}. Regions II ($u>0,$ $v<0$) and III ($%
u<0,$ $v>0$) are occupied by the approaching waves with line element (1).
Region I ($u<0,$ $v<0$) represents the flat spacetime between the waves.
Region IV ($u>0,$ $v>0$) describes their collision and interaction. We
suppose that in the interaction region the line element preserves its axial
symmetry and is described by functions $g_{uv}=e^{-M},$ $U$ and $V$ which
depend on $u,$ $v,$ $r.$ In the present paper we shall find all solutions
with $M=0$, in a manner similar to the classification of diagonal plane
waves with $M=0$ \cite{seven}.

In Sec.II the general solution with $M=0$ is found in the interaction
region. It is extended to a global solution in Sec.III and its parameters
are linked to the characteristics of the approaching waves. The structure of
the solution is elucidated further in Sec.IV by studying its invariants. In
Sec.V two examples are given. Sec.VI contains some conclusions.

\section{Solution with $M=0$ in the interaction region}

The vacuum Einstein equations in the interaction region simplify when (1) is
rewritten as 
\begin{equation}
ds^2=2dudv-Q^2dr^2-P^2d\varphi ^2  \label{two}
\end{equation}
Then they read: 
\begin{equation}
PQ_{uu}+QP_{uu}=0  \label{three}
\end{equation}

\begin{equation}
PQ_{vv}+QP_{vv}=0  \label{four}
\end{equation}
\begin{equation}
Q_{uv}=0  \label{five}
\end{equation}
\begin{equation}
P_{uv}=0  \label{six}
\end{equation}
\begin{equation}
Q_vP_r=QP_{vr}  \label{seven}
\end{equation}
\begin{equation}
Q_uP_r=QP_{ur}  \label{eight}
\end{equation}
\begin{equation}
Q^2\left( P_vQ_u+P_uQ_v\right) -QP_{rr}+P_rQ_r=0  \label{nine}
\end{equation}
When $P_r=Q_r=0$ eqs(7,8) become trivial and the others reduce to the
equations for plane waves \cite{seven}.

Eqs(7,8) are easily integrated and give $Q=e^{h\left( r\right) }P_r$ with an
arbitrary $h\left( r\right) $. However, when $u=v=0$ we are in region I with
Minkowskian background and $P\left( 0,0,r\right) =r,$ $Q\left( 0,0,r\right)
=1$. This condition fixes $h\left( r\right) $ to zero and the result is 
\begin{equation}
Q=P_r  \label{ten}
\end{equation}
coinciding with the condition for a single $pp$-wave \cite{six}. In fact,
the final conclusion drawn below does not depend on $h\left( r\right) $. Now
(3,4) become 
\begin{equation}
\left( P_{uu}P\right) _r=0  \label{eleven}
\end{equation}
\begin{equation}
\left( P_{vv}P\right) _r=0  \label{twelve}
\end{equation}
Eqs(5,6) show that the $u$ and $v$ dependence separate: 
\begin{equation}
P=f\left( u,r\right) +g\left( v,r\right)  \label{thirteen}
\end{equation}
The remaining eq(9) may be written in two equivalent forms: 
\begin{equation}
\left( e^{-U}\right) _{uv}=0  \label{fourteen}
\end{equation}
\begin{equation}
\left( P_uP_v\right) _r=0  \label{fifteen}
\end{equation}
The first form is well-known from the study of colliding plane waves.

We want to prove that the solution of (11-15) is 
\begin{equation}
P=b_0\left( r\right) -b_1\left( r\right) u-b_2\left( r\right) v
\label{sixteen}
\end{equation}
for some functions $b_i\left( r\right) $. If $P_{uu}=P_{vv}=0$ eq(16)
immediately follows. Suppose that $P_{vv}\neq 0$. Then (11,12) and (13-15)
give 
\begin{equation}
\left( \frac{P_{uu}}{P_{vv}}\right) _r=0  \label{seventeen}
\end{equation}
\begin{equation}
\left( P_{uu}P_{vv}\right) _r=0  \label{eighteen}
\end{equation}
which are equivalent to 
\begin{equation}
P_{uur}P_{vv}=0  \label{nineteen}
\end{equation}
\begin{equation}
P_{uu}P_{vvr}=0  \label{twenty}
\end{equation}

There are two possibilities: $P_{uur}=P_{vvr}=0,$ $\left( P_{uu}\neq
0\right) $ or $P_{uu}=0$. The first possibility combined with (13) means
that $P_r=0$. Hence (3-9) reduce to the plane wave case, as was already
mentioned, which is discussed in \cite{seven}. The second possibility means
that $f=c_1\left( r\right) u+c_2\left( r\right) $. Putting this result into
(13-15) we get $c_1\left( r\right) g_v=g_1\left( v\right) $. Eqs(11,12)
become 
\begin{equation}
g_{1v}\left( u+\frac{c_2+g}{c_1}\right) _r=0  \label{twentyone}
\end{equation}
If $g_{1v}=0$ it easily follows that $P$ takes the form (16). If $g_{1v}\neq
0$, $g=c_1g_2\left( v\right) -c_2$ and (17,18) give $\left(
c_1^2g_{2v}\right) _r=0$. Again, $P$ is of the form (16).

Suppose, at last, that $P_{vv}=0$ but $P_{uu}\neq 0$. Since the equations
are symmetric with respect to $u,$ $v$ the same argument leads to the same
conclusion. Inserting (16) into (15) we obtain the constraint 
\begin{equation}
b_1\left( r\right) b_2\left( r\right) =a  \label{twentytwo}
\end{equation}
where $a\neq 0$ is some constant.

\section{Global solution with M=0}

It is obtained by taking into account the Minkowski boundary condition and
extending the solution from region IV to regions II and III with the help of
the Penrose ansatz: 
\begin{equation}
P=r-b_1\left( r\right) u\theta \left( u\right) -b_2\left( r\right) v\theta
\left( v\right)  \label{twentythree}
\end{equation}
\begin{equation}
Q=1-b_{1r}u\theta \left( u\right) -b_{2r}v\theta \left( v\right)
\label{twentyfour}
\end{equation}

Going to region II or region III we see that the approaching waves are
impulsive \cite{six}: 
\begin{equation}
b_i\left( r\right) =H_i\left( r\right) _r  \label{twentyfive}
\end{equation}
\begin{equation}
ds^2=2du_idw_i+2H_i\left( P\right) \delta \left( u_i\right)
du_i^2-dP^2-P^2d\varphi ^2  \label{twentysix}
\end{equation}
where $u_1=u,$ $u_2=v$. They are induced by some impulse with energy-density 
\begin{equation}
\rho _i\left( r,u_i\right) =\frac 1{2r}\left( rH_i\left( r\right) _r\right)
_r\delta \left( u_i\right)  \label{twentyseven}
\end{equation}
due to a beam of pure radiation \cite{eight}, light \cite{nine} or a
point-particle moving with the speed of light \cite{ten}. Then (22) becomes 
\begin{equation}
H_2\left( r\right) _r=\frac a{H_1\left( r\right) _r}  \label{twentyeight}
\end{equation}
It is clear that (28) prevents the study of two equal colliding waves, e.g.
two ultrarelativistic black holes. This is a consequence of the simplifying
assumption $M=0$. Positive energy-density induces positive and increasing $%
H_i$, hence $b_i>0,$ $b_{ir}>0,$ $a>0$. Applying the constraint (22) to
(23,24) and changing notation to $b_1\equiv b$ yields 
\begin{equation}
P=r-bu\theta \left( u\right) -\frac abv\theta \left( v\right)
\label{twentynine}
\end{equation}

\begin{equation}
Q=1-b_ru\theta \left( u\right) +\frac{ab_r}{b^2}v\theta \left( v\right)
\label{thirty}
\end{equation}

The change of the relative sign in $Q$ is reminiscent of the similar change
in the Babala solution \cite{eleven} which is one of the three diagonal
vacuum plane waves with $M=0$ \cite{seven}. Thus (29,30) may be considered
as a one-function analog of the Babala solution, although they do not reduce
to it when $b$ is constant. Eqs(29,30) also give 
\begin{equation}
e^{-U}=r-\left( rb_r+b-bb_ru\right) u\theta \left( u\right) +\frac a{b^2}%
\left( rb_r-b-\frac{ab_r}bv\right) v\theta \left( v\right) 
\label{thirtyone}
\end{equation}
The presence of null matter at the boundaries is signalled in the
coordinates (1) by the discontinuities in $U_u$ which break the
O'Brien-Synge boundary conditions \cite{one}. The terms linear in $u$ and $v$
disappear from (31) when $rb_r\pm b=0$ and these conditions are satisfied
simultaneously only by a trivial $b$.

\section{Structure of the invariants}

More information about the solution may be learned from its invariants. The
only non-trivial Ricci scalars are 
\begin{equation}
\Phi _{22}=\frac 12R_{uu}=-\frac 12e^U\left( P_{uu}P\right) _r=re^U\rho
_1\left( r,u\right)  \label{thirtytwo}
\end{equation}
\begin{equation}
\Phi _{00}=\frac 12R_{vv}=-\frac 12e^U\left( P_{vv}P\right) _r=re^U\rho
_2\left( r,v\right)  \label{thirtythree}
\end{equation}
from which one can deduce the energy-momentum tensor: 
\begin{equation}
T_{\mu \nu }=2re^U\left( \rho _1l_\mu l_\nu +\rho _2n_\mu n_\nu \right)
\label{thirtyfour}
\end{equation}
where $l_\mu ,$ $n_\mu $ are the first two vectors of the usual NP tetrad
for (2) \cite{one}. There are two planes of null dust with variable energy
densities. In regions II, III they coincide with (27) but along the
boundaries of IV they become dependent on the other null coordinate because $%
re^U\rho _i\neq \rho _i$. The factor $e^U$ in (32-34) is well-known in plane
wave solutions with thin shells of null-matter \cite{eleven}, \cite{twelve}.
Eq(29) shows that in regions II and III the single $pp$-waves have
coordinate singularities $P=0$. Eqs(32,33) tell that they turn into
curvature singularities on the boundaries of the interaction region at
points $v=0,$ $u=\frac rb$ and $u=0,$ $v=\frac{rb}a$.

The only non-trivial Weyl scalars are 
\begin{equation}
\Psi _4=\frac{P_{uu}}P+\Phi _{22}=-\frac{b^2\delta \left( u\right) }{%
rb-av\theta \left( v\right) }+\Phi _{22}  \label{thirtyfive}
\end{equation}
\begin{equation}
\Psi _0=\frac{P_{vv}}P+\Phi _{00}=-\frac{a\delta \left( v\right) }{b\left(
r-bu\theta \left( u\right) \right) }+\Phi _{00}  \label{thirtysix}
\end{equation}
From (27,32,33,35,36) it is clear that the interaction region is conformally
flat. The Weyl scalars confirm that the approaching waves are impulsive.
They, like the Ricci scalars, become singular when $P$ vanishes. These point
curvature singularities are generic and cannot be avoided by a careful
choice of $b$. In regions II, III eqs(35,36) coincide with the expressions
derived in \cite{six}.

\section{Some examples}

The energy-densities may be given as functions of $b$ by (22,25,27): 
\begin{equation}
\rho _1=\frac{\left( rb\right) _r}{2r}\delta \left( u\right) 
\label{thirtyseven}
\end{equation}
\begin{equation}
\rho _2=\frac a{2r}\left( \frac rb\right) _r\delta \left( v\right) 
\label{thirtyeight}
\end{equation}
The second density is positive when $\frac rb$ is an increasing function.
One possible solution includes an ultrarelativistic black hole approaching
from region II and is given by 
\begin{equation}
b=\frac{4\mu }r  \label{thirtynine}
\end{equation}
\begin{equation}
H_1=4\mu \ln r  \label{forty}
\end{equation}
\begin{equation}
H_2=\frac{ar^2}{8\mu }  \label{fortyone}
\end{equation}
\begin{equation}
\rho _1=\frac \mu 2\delta \left( r\right) \delta \left( u\right) 
\label{fortytwo}
\end{equation}
\begin{equation}
\rho _2=\frac a{4\mu }\delta \left( v\right)   \label{fortythree}
\end{equation}
\begin{equation}
P=r-\frac{4\mu }ru\theta \left( u\right) -\frac{ar}{4\mu }v\theta \left(
v\right)   \label{fortyfour}
\end{equation}
\begin{equation}
Q=1+\frac{4\mu }{r^2}u\theta \left( u\right) -\frac a{4\mu }v\theta \left(
v\right)   \label{fortyfive}
\end{equation}
\begin{equation}
\Phi _{00}=\frac{ar^4\delta \left( v\right) }{4\mu \left( r^4-16\mu
^2u^2\theta \left( u\right) \right) }  \label{fortysix}
\end{equation}
\begin{equation}
\Phi _{22}=\frac{8\mu ^3\delta \left( r\right) \delta \left( u\right) }{%
\left( 4\mu -av\theta \left( v\right) \right) ^2}  \label{fortyseven}
\end{equation}
\begin{equation}
\Psi _0=-\frac{ar^2u\theta \left( u\right) \delta \left( v\right) }{%
r^4-16\mu ^2u^2\theta \left( u\right) }  \label{fortyeight}
\end{equation}
\begin{equation}
\Psi _4=\frac{8\mu ^2\delta \left( u\right) }{4\mu -av\theta \left( v\right) 
}\left( \frac{\mu \delta \left( r\right) }{4\mu -av\theta \left( v\right) }-%
\frac 2{r^2}\right)   \label{fortynine}
\end{equation}
where $\mu $ is the momentum of the null point-particle. The wave arriving
from region III is induced by a pure radiation burst of constant density
across the wavefront. In fact, this is a plane wave \cite{six}. Speaking
loosely, this example describes the collision of an almost pure exterior
solution with the simplest interior solution.

There is another solution with positive $\rho _i$ which are finite on the
axis $r=0$ and decrease when $r\rightarrow \infty $. It is given by 
\begin{equation}
b=\frac 1r\ln \left( c+r^2\right)   \label{fifty}
\end{equation}
\begin{equation}
\rho _1=\frac 1{c+r^2}\delta \left( u\right)   \label{fiftyone}
\end{equation}
\begin{equation}
\rho _2=\frac a{\ln \left( c+r^2\right) }\left[ 1-\frac{r^2}{\left(
c+r^2\right) \ln \left( c+r^2\right) }\right] \delta \left( v\right) 
\label{fiftytwo}
\end{equation}
with $c>e$. One can say that the waves are beam-like i.e. they have finite
transverse extent, but their energy diverges. It seems impossible to arrange
for finite energy of both waves when $M=0$. We omit the lengthy expressions
for the metric and its invariants because the solution possesses the general
features established above.

\section{Conclusion}

The assumption $M=0$ in the case of colliding axisymmetric $pp$-waves is
almost as restrictive as in the case of colliding plane waves, although the
freedom in the solution extends to an arbitrary function $b$ instead of
arbitrary constants. The solution in the interaction region is still
conformally flat and linear in $u$ and $v$. This indicates the presence of
null matter along the boundaries, but anyway for $pp$-waves the distinction
between pure gravitational and matter field components is not so clean-cut
in view of relations like (27). The mechanism by which the coordinate
singularities turn into curvature singularities is the same as for plane
waves and has its roots in (14). The constraint (22) does not allow to study
the simplest possible case of two equivalent approaching waves. Obviously,
more exact solutions are necessary and non-trivial interactions between $pp$%
-waves of finite energy should be possible when the condition $M=0$ is
relaxed.

{\bf Acknowledgements}

This work was supported by the Bulgarian National Fund for Scientific
Research under contract F-632.

\end{document}